\documentclass[global,twocolumn]{svjour}

\usepackage{amsmath,amssymb}
\usepackage{graphics}

\journalname{Appl. Phys. B}

\begin{document}

\title{Ultra-cold atoms in an optical cavity: Two-mode laser locking to the cavity avoiding radiation pressure}
\author{Simone Bux \and Gordon Krenz \and Sebastian Slama \and Claus Zimmermann \and Philippe W. Courteille}

\institute{Physikalisches Institut, Eberhard-Karls-Universit\"at T\"ubingen,
\\Auf der Morgenstelle 14, D-72076 T\"ubingen, Germany}

\date{Received: date / Revised version: date}

\maketitle

\begin{abstract}
The combination of ultra-cold atomic clouds with the light fields of optical cavities provides a powerful model system for the development of new types of laser cooling and for studying cooperative phenomena. These experiments critically depend on the precise tuning of an incident pump laser with respect to a cavity resonance. Here, we present a simple and reliable experimental tuning scheme based on a two-mode laser spectrometer. The scheme uses a first laser for probing higher-order transversal modes of the cavity having an intensity minimum near the cavity's optical axis, where the atoms are confined by a magnetic trap. In this way the cavity resonance is observed without exposing the atoms to unwanted radiation pressure. A second laser, which is phase-locked to the first one and tuned close to a fundamental cavity mode drives the coherent atom-field dynamics.
\end{abstract}


\maketitle

\section{Introduction}


Optical cavities have been proposed for efficient cooling of atoms and molecules to ultra-low temperatures \cite{Gangl00,Vuletic00,Domokos03}. In contrast to conventional Doppler cooling \cite{Haensch75}, where the particles' excess kinetic energy is dissipated via spontaneous decay of internal atomic excitation, cavity cooling also works when the pump laser light is tuned very far from an atomic resonance. Dissipation is provided by the cavity mirrors whose finite transmission reduces the lifetime of the optical modes. Hence, the cooling limit is ultimately set by the cavity linewidth and can be far below the Doppler-cooling limit if the cavity's finesse is high. Since cavity-cooling works far away from the atomic resonances, it suits well for molecules whose level schemes do not lend themselves to standard cooling methods. An important issue is that the cavity cooling power depends dispersively on the detuning of the pump laser from the \textit{cavity resonance}. Only when the pump light is detuned from a cavity mode, the cavity losses take away excess kinetic energy of atoms stored in the cavity field.

Another application of optical cavities is the study of collective \cite{Kruse03b,Nagorny03b} and self-organization phenomena \cite{Black03,Cube04} observed in the coupled dynamics of ultra-cold atoms interacting with the cavity modes. The technical development culminated with the recent realization of Bose-Einstein condensation in a high-finesse ring cavity \cite{Slama07,Slama07b}. For these experiments, a precise timing of the pump laser irradiation and the ability to choose the pump laser power and detuning over wide ranges is often essential. In particular, some subtle effects in the collective dynamics only emerge at very low pump laser powers \cite{Piovella01b}.

To satisfy the requirements of tunability, fast switching and variable pump power, it is necessary to reference the laser with respect to the cavity's mode structure. The most convenient way is to use \textit{two} lasers: One laser (which we will call reference laser) serves to probe the cavity's free spectral range. The cavity's response to the injected reference light field can then be used to set up a servo loop. The other laser (called pump laser) is locked to the reference laser via a phase-locked loop (PLL). The pump laser drives the collective atomic dynamics. Its frequency can be tuned with high precision by tuning the local oscillator of the PLL, and its power can readily be switched on and off without perturbing the reference laser lock to the cavity.

For any cavity servo lock to operate reliably a mini\-mum of injected light power is required. This unfortunately results in radiation pressure, which heats the atomic cloud and reduces the efficiency of cavity cooling or perturbs the collective dynamics. The question is therefore, how to probe a cavity without subjecting the atoms to radiation pressure. One solution is to tune the reference laser very far from any atomic resonance, where radiation pressure is weak. If the pump laser is to be operated near an atomic resonance, the two lasers will have very different frequencies. They can be referenced to each other e.g.~via a transfer cavity \cite{Helmcke87}, using frequency mixing techniques \cite{Telle90} or frequency combs \cite{Holzwarth00}. The disadvantage of those approaches is that they all involve many servo loops (typically up to 5) to work simultaneously. Furthermore, the construction of a stable transfer cavity or a frequency comb is expensive.

The request that the reference laser does not interact with the atoms can be met by an alternative approach, which consists in running the reference laser on a higher-order transversal mode TEM$_{kl}$ of the cavity. Atoms can be stored in higher-order TEM$_{kl}$ modes as has been demonstrated earlier \cite{Kruse03}. This is however not the purpose of the present work. On the contrary, we now want to avoid exposition of the atoms to the TEM$_{kl}$ modes by placing them within the dark zones in between the high intensity regions by means of a magnetic trap. The advantage of this approach is that the reference laser can be tuned close to the pump laser so that their frequencies can be locked via a single phase-locked loop.

We begin this paper with a presentation of the experimental setup and then calculate the scattering and heating rate of an ultra-cold atomic cloud stored in a TEM$_{kl}$ mode close to the optical axis of the cavity. To demonstrate that the realized scheme yields sufficiently low scattering rates, we measure the atomic lifetime and the heating rate. Finally, we characterize our two-mode laser locking scheme.

\section{Experimental setup}
\label{SecSetup}

The details of our experimental apparatus have been presented in previous publications \cite{Slama07,Slama07b}. A standard magneto-optical trap located in an ultra-high vacuum chamber is loaded with rubidium atoms, which are subsequently transferred into a magnetic quadrupole trap and then compressed into a Ioffe-type trap generated by four parallel wires. The atoms are cooled by forced evaporation down to a temperature of a few micro-Kelvin or even to quantum degeneracy. Finally, the magnetic trap containing typically $N\simeq2\times10^6$ ultra-cold atoms is shifted into the mode volume of a laser-driven high-finesse ring cavity. At this location the radial and axial secular frequencies of the magnetic trap are $\omega_{\rho}=(2\pi)~180~$Hz and $\omega_z=(2\pi)~40~$Hz, respectively.

\begin{figure}[ht]
		\centerline{\scalebox{0.7}{\includegraphics{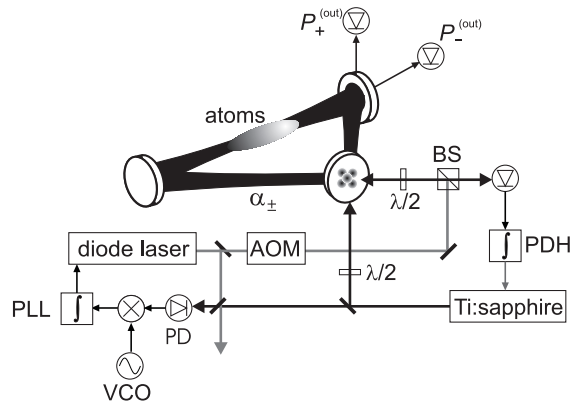}}}
		\caption{Principle scheme of the optical setup. A Ti:sapphire laser is locked via a PDH servo to a TEM$_{11}$ mode of the ring
			cavity. A diode laser is locked via a PLL to the Ti:sapphire laser and injected into a counterpropagating TEM$_{00}$ ring 
			cavity mode. The diode laser beam can be shuttered with an acousto-optic modulator. Via $\lambda/2$ waveplates the 
			polarization of the light coupled into the cavity can be switched between $s$- and $p$-polarization. BS: non-polarizing beam
			splitter, PD: fast photodetector.}
		\label{Fig1}
	\end{figure}

The decay rate of the ring cavity is for $s$-polarized light $\kappa_c=(2\pi)~270~$kHz, which corresponds to a finesse of $F=6400$. It can be changed to $\kappa_c=(2\pi)~20~$kHz ($F=87000$) by injecting $p$-polarized light. The free spectral range is $\delta_{fsr}=3.4~$GHz and the waist of the cavity mode is $w_0=107~\mu$m. The cavity is driven by two lasers (see Fig.~\ref{Fig1}) coupled in from different directions. The reference laser, a home-built Ti:sapphire laser, is stabilized to a higher-order transversal mode of the ring cavity (in general TEM$_{10}$ or TEM$_{11}$) using the Pound-Drever-Hall (PDH) technique. The servo bandwidth is about $200~$kHz \cite{Kruse03b}. The fast components of the error signal are fed to an acousto-optic modulator (AOM) not shown in Fig.~\ref{Fig1}. The pump laser is a diode laser, which is offset-locked to the Ti:sapphire laser by means of a PLL \cite{Prevedelli95,Schunemann99}. To this end, both laser beams are phase-matched on a fast photodiode (12~GHz bandwidth). The beat signal is mixed with the output of a stable voltage controlled oscillator (VCO) and, via a loop filter, directly fed back to the laser diode current. The diode laser is injected into a TEM$_{00}$ mode of the cavity, whose frequency can be detuned as far as several giga-Hertz from the TEM$_{11}$ mode. An important advantage of this scheme, which only consists of two servo loops, is that the pump light can be switched on and off rapidly with an AOM without interfering with the PDH servo.

\section{Ultra-cold atoms in a TEM$_{kl}$ mode}

In the following we will show that the injection of a pump laser with the purpose of driving collective instabilities in an atomic cloud in the ring cavity is intrinsically connected with radiation pressure. The atom-field coupling strength is measured by a quantity called collective gain \cite{Slama07,Slama07b,Piovella01b}, $G\simeq nNU_1^2/\kappa_c$. The photon number $n$ in the pump laser mode is related to the intracavity laser power $P=n\hbar\omega\delta_{fsr}$. $U_1=g_1^2/\Delta_a$ is the single-photon light shift, where $\Delta_a$ is the detuning of the pump laser from an atomic resonance, e.g.~the rubidium $D_1$ line, and the single-photon Rabi frequency is $g_1=\sqrt{3\Gamma\delta_{fsr}}/kw_0\simeq(2\pi)~80~$kHz for our ring cavity. The Rayleigh scattering rate can roughly be estimated by $R\simeq N\sigma I/\hbar\omega$, where $I=2P/\pi w_0^2$ is the light intensity. Far from resonance the optical cross section can be approximated by $\sigma=3\lambda^2\Gamma^2/8\pi\Delta_a^2$. From this follows that the scattering rate is equivalent to the gain divided by the cooperativity parameter $g_1^2/\Gamma\kappa_c$ characterizing the cavity,
\begin{equation}\label{Eq01}
	R=G\frac{\Gamma\kappa_c}{g_1^2}~.
\end{equation}
Hence, for a given design of the cavity, a wanted gain $G$ is always accompanied by a Rayleigh scattering rate $R$. Note that the pump laser driving the collective dynamics is tuned close to a TEM$_{00}$ mode of the ring cavity. While we can not impede radiation pressure exerted by the (weak) pump laser, we can at least avoid additional radiation pressure by the reference laser, which should not participate in the collective dynamics. I.e.~the reference laser must generate lower scattering rates than the pump laser. In the same time the reference laser power must be sufficiently strong to guarantee stable operation of the PDH servo. We achieve this by tightly phase-locking this laser to a TEM$_{11}$ mode. If the atomic cloud is sufficiently small, it may be contained within a region of space where the intensity of the TEM$_{11}$ mode is negligibly small, so that radiation pressure is efficiently reduced. 

\bigskip

\begin{figure}[ht]
		\centerline{\scalebox{0.45}{\includegraphics{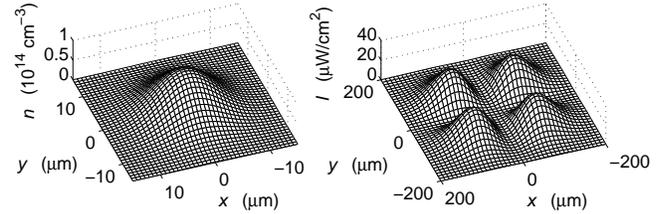}}}\caption{
			(a) Radial density distribution of a $T=1~\mu$K cold cloud of $N=10^5$ atoms.
			(b) Radial intensity profile of the TEM$_{11}$ mode for a total power of $P=10~$mW.}
		\label{Fig2}
	\end{figure}

The scattering rate off an atomic cloud exposed to a higher-order TEM$_{kl}$ mode of the ring cavity can be estimated from the overlap of the intensity distribution $I_{kl}$ with the atomic density distribution $n$ (see Fig.~\ref{Fig2}),
\begin{equation}\label{Eq02}
	R_{kl}=\int\sigma\frac{I_{kl}(\mathbf{r})}{\hbar\omega}n(\mathbf{r})d^3\mathbf{r}~.
\end{equation}
The intracavity intensity distribution is \cite{Kogelnik66}
\begin{equation}\label{Eq03}
	I_{kl}(\mathbf{r})=I_0e^{-2\rho^2/w^2}H_k(\sqrt{2}x/w)^2H_l(\sqrt{2}y/w)^2~,
\end{equation}
where $\rho^2\equiv x^2+y^2$ and $H_k(\xi)$ are the Hermite polynomials. The constant $I_0$ is determined from the normalization condition $P=\int I_{kl}(\mathbf{r})dxdy=\pi w^2I_0/2$. It is a good assumption that the atomic cloud is located within the Rayleigh length, so that the rms-beam diameter is $w(z)\approx w_0$. Furthermore, if the atomic cloud is located very close to the optical axis, we can approximate $e^{-2\rho^2/w^2}\approx1$. Restricting to $k,l\leq1$, the Hermite polynomials take a simple form,
\begin{equation}\label{Eq04}
	I_{kl}(\mathbf{r})\approx I_0(\sqrt{2}x/w)^{2k}(\sqrt{2}y/w)^{2l}~.
\end{equation}

\bigskip

The atomic density distribution of a trapped thermal gas is given by $n(\mathbf{r})=n_0e^{-U(\mathbf{r})/k_BT}$. When the trapping potential is harmonically approximated using the radial and axial secular frequencies, the integral~(\ref{Eq02}) is easily solvable,
\begin{equation}\label{Eq05}
	R_{kl}=\frac{3N\Gamma^2P}{k^2w^2\Delta_a^2\hbar\omega}\frac{2^{3/2}\bar{\rho}}{\bar{z}}\left(\frac{2\bar{\rho}}{w}\right)^{2k+2l}~,
\end{equation}
where $\bar{\rho}\equiv\sqrt{k_BT/m\omega_{\rho}^2}$ and $\bar{z}\equiv\sqrt{k_BT/m\omega_z^2}$ are the rms-radii of the atomic cloud. The overlap of the atomic cloud with the mode's intensity distribution dramatically diminishes as the cloud's rms-radius shrinks. At ultra-low temperatures the rms-radius is much smaller than the beam waist, as shown in Fig.~\ref{Fig2}.

This results in an improvement of the scattering rate which can be expressed as
\begin{equation}\label{Eq06}
	\frac{R_{11}}{R_{01}}=\frac{R_{01}}{R_{00}}=\frac{4k_BT}{m\omega_{\rho}^2w^2}~.
\end{equation}
Hence, the ratio $R_{11}/R_{00}$ depends quadratically on the temperature. At $T=1~\mu$K it takes the value $R_{11}/R_{00}\approx10^{-4}$. 

\bigskip

Rayleigh scattering heats up the atomic cloud and causes trap losses. If the atoms are Bose-Einstein condensed, every scattering event expels an atom from the condensate. The heating rate is given by \cite{Grimm00}
\begin{equation}\label{Eq07}
	k_B\dot{T}=\frac{1}{3}\hbar\omega_rR_{kl}~,
\end{equation}
where $\omega_r=\hbar k^2/m$ is the recoil shift. The heating is appropriately described by Eq.~(\ref{Eq07}) only if the cloud rapidly thermalizes. The collision rate estimated for our experiments, $\gamma_{coll}=20~$s$^{-1}$, is large enough to satisfy this condition.

A second trap loss mechanism is due to spin relaxation which redistributes the population of the magnetically trapped $|F=2,m_F=2\rangle$ state among all substates of the hyperfine multiplet including those which are expelled from the magnetic trap. Therefore, spin relaxation is a caused by Raman scattering processes. The ratio between Raman and Rayleigh scattering is given by \cite{Cline94}
\begin{equation}\label{Eq08}
	\frac{R_{Ram}}{R_{Ray}}=\left(\frac{\Delta_{D1}-\Delta_{D2}}{2\Delta_{D1}+\Delta_{D2}}\right)^2~,
\end{equation}
where $\Delta_{D1,2}$ are the laser detunings from the $D_1$ and $D_2$ lines. Hence, spin relaxation is particularly strong near the atomic fine structure, when the difference between the detunings $\Delta_{D1}$ and $\Delta_{D2}$ can not be neglected. For example, at $797~$nm we expect $\frac{R_{Ram}}{R_{Ray}}\simeq 50\%$.

\section{Measurements of scattering rates}

To measure the lifetime of the atomic cloud in the presence of light in the cavity, we proceed as follows. The reference laser light is matched and continuously phase-locked to a specific TEM mode of the ring cavity. The atomic cloud is trapped and cooled slightly above the mode volume. When the end of the evaporation ramp is reached, the magnetic trap center is lowered within $100~$ms to coincide with the optical axis of the cavity. Here, the atom cloud is held for a variable time period, before the magnetic trapping fields and the reference laser beam are simultaneously and rapidly shut down. Finally, the atomic cloud is absorption-imaged after a time of ballistic expansion of $15~$ms. The number of atoms remaining in the trap after this time and the temperature increase are recorded as a function of the holding time.

\bigskip

\begin{figure}[ht]
		\centerline{\scalebox{0.48}{\includegraphics{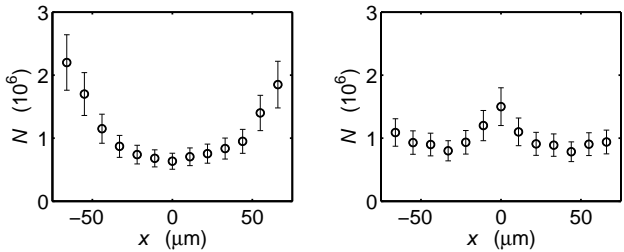}}}\caption{
			Variation of the lifetime as the horizontal position of the atomic cloud is varied for a TEM$_{00}$ mode (a) and a
			TEM$_{10}$ mode (b). $N$ is the number of atoms remaining in the magnetic trap due to heating and/or spin
			relaxation after $1~$s interaction time.
			The errorbars include statistical and systematic errors. }
		\label{Fig3}
	\end{figure}

Our method to optimize the atomic position consists in measuring the Rayleigh scattering rate, which depends on the precise matching of the atomic cloud's position with respect to the optical mode. Therefore, upon shifting the magnetic trap along one direction, e.g.~the horizontal $x$-direction, for a TEM$_{00}$ mode we expect maximum scattering when the position is centered. In contrast, for a TEM$_{10}$ mode the scattering rate should have a minimum for $x=0$, and increase when the cloud is displaced. This is seen in Fig.~\ref{Fig3}.

\bigskip

We carried out measurements of trap losses and heating rates after variable interaction times between the cavity mode and the atomic cloud. The atomic losses are only partly due to heating. A Stern-Gerlach type experiment reveals a noticeable population of the trapped $|F=2,m_F=1\rangle$ state that results from light-induced spin relaxation of atoms out of the $|F=2,m_F=2\rangle$ state. This emphasizes the important role of Raman transitions.

\begin{figure}[ht]
		\centerline{\scalebox{0.48}{\includegraphics{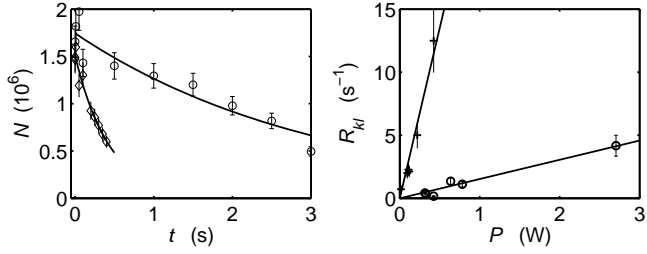}}}
		\caption{(a) Decrease in atom numbers due to Rayleigh scattering for the TEM$_{00}$ mode (diamonds) and the TEM$_{11}$ 
			mode (circles). The laser powers are $P=58~$mW for the TEM$_{00}$ and $P=174~$mW for the TEM$_{11}$, the wavelength is $\lambda=797.7~$nm in both cases. The initial temperature is $T=3~\mu$K.
			(b) Rayleigh scattering rates for the TEM$_{00}$ (crosses) and the TEM$_{11}$ modes (circles) as a function of laser power. 
			The data points have been taken at different laser detunings $\Delta_a$. In order to allow for a direct comparison, the
			scattering rates have been normalized to a detuning $\Delta_{ar}$ corresponding to a wavelength of $\lambda_r=798~$nm 
			via $R_{kl}\rightarrow\Delta_a^2R_{kl}/\Delta_{ar}^2$.}
		\label{Fig4}
	\end{figure}

Fig.~\ref{Fig4}(a) demonstrates that the lifetime of an atomic cloud cooled to a temperature of $T=3~\mu$K is dramatically improved in the TEM$_{11}$ mode as compared to the TEM$_{00}$ mode. The corresponding curves have been recorded under different conditions, i.e. different laser detunings $\Delta_{a,kl}$ and powers $P_{kl}$. The lifetime enhancement is thus $R_{00}\Delta_{a,00}^2P_{11}/R_{11}\Delta_{a,11}^2P_{00}\simeq20$. This is also seen in Fig.~\ref{Fig4}(b), which exhibits the power dependence of the measured Rayleigh scattering rates. In order to compare the scattering rates at a chosen detuning $\Delta_{ar}$ we rescale them via $R_{kl}\rightarrow\Delta_a^2R_{kl}/\Delta_{ar}^2$. The power-dependence of the $R_{00}$ is 20 times steeper than that of $R_{11}$. The lifetime enhancement is considerably less than the theoretical 1000-fold improvement predicted by Eq.~(\ref{Eq06}) for a cloud at $T=3~\mu$K. The main reason for this discrepancy are uncertainties in the exact positioning of the atomic cloud into the center of the TEM$_{11}$ cavity mode. Furthermore, during the transfer of the atoms from their initial location to the center of the cavity mode the atoms may traverse zones of high light intensity. The resulting increase in the spatial overlap between the light intensity distribution and the atomic density distribution leads to enhanced scattering. Based on Eq.~(\ref{Eq02}) we calculate for a displacement of the atoms from the cavity mode center of only $13~\mu$m in $x$ and $y$ direction a 50-fold enhancement of the scattering rate, which would account for the discrepancy. Such a displacement is below the resolution of our calibration measurement used to center the cloud shown (see Fig.~\ref{Fig3}). A second reason for enhanced scattering could be diffraction of the edges of the light mode by the Ioffe wires generating the magnetic trapping potential. This results in stray light capable of heating the atomic cloud. We however verified that this effect is small enough to be neglected even for higher-order TEM modes, which radially extend further into space.
	
\begin{figure}[ht]
		\centerline{\scalebox{0.48}{\includegraphics{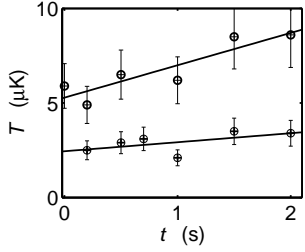}}}
		\caption{Temperature evolution for magnetically trapped atoms without light in the cavity (upper curve) and in the 
			TEM$_{11}$ mode (lower curve). While heating of about $1.7~\mu$K/s is observed without light, pumping the TEM$_{11}$ 
			mode leads to a reduction of the heating rate to about $0.5~\mu$K/s. The solid lines are linear fits to the data points. Here, the laser power is $P=116~$mW and the laser wavelength is $\lambda=796.4~$nm. Note that the initial temperature is different for both cases.}
		\label{Fig5}
	\end{figure}

Figure~\ref{Fig5} shows the evolution of the temperature $T$ as a function of the interaction time. Interestingly the heating rate decreases when the atoms are placed in the TEM$_{11}$ mode. This is due to the inhomogeneity of the Rayleigh scattering. Because only the outer regions of the atomic cloud are exposed to light, only the hottest atoms are heated and lost from the trap. Since the estimated collision rate of $20~\text{s}^{-1}$ is sufficiently high to ensure rapid thermalization of the remaining atoms, the heating rate is effectively reduced. As shown in Fig.~\ref{Fig5}, noise in the trapping potentials heats up the atomic cloud for long storage times, even if there is no light in the cavity. The heating rate is lower if the TEM$_{11}$ mode is optically pumped.

\section{Performance of the locking scheme}

Previous experiments \cite{Slama07} performed with the ring cavity have been affected by the finite rise time of the pump mode. This time was limited by the response time of the PDH servo loop. With the new setup the pump laser can be switched using an acousto-optic modulator without perturbing the locking servo. Hence, the only limit to the rise time is the cavity decay time, which is $4~\mu$s for high-finesse and $200~$ns for low finesse.

\begin{figure}[h]
		\centerline{\scalebox{0.48}{\includegraphics{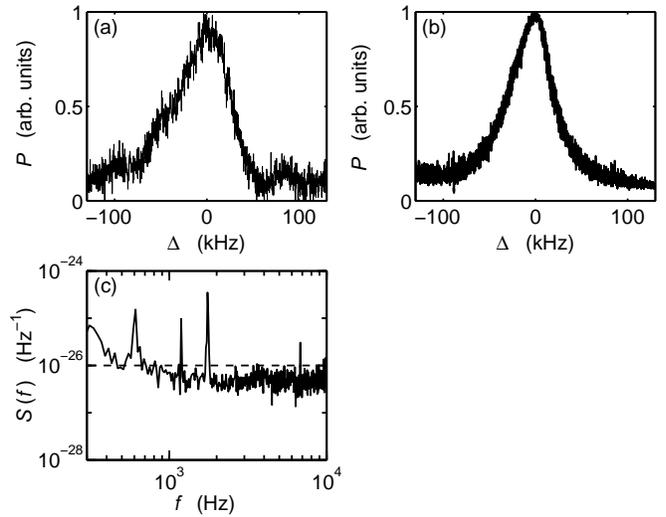}}}
		\caption{(a) Profile of the transmitted intensity $P$ of a TEM$_{00}$ mode recorded in high-finesse with the diode laser. 
			The reference laser was locked in high finesse on a TEM$_{11}$ mode. The asymmetric lineshape is explained in the text.
			(b) Transmission profile of a TEM$_{04}$ mode in high-finesse. The spectral width (FWHM) is $41~$kHz, which coincides 
			with the calculated value of $40~$kHz. Here the Ti:sapphire laser is locked to a TEM$_{00}$ mode in high-finesse.
			(c) Spectral density of normalized frequency noise. The dashed line draws a white noise spectrum corresponding to 
			a laser with 2~kHz emission bandwidth.}
		\label{Fig6}
	\end{figure}

To characterize the overall accuracy of the realized two-mode laser spectrometer, we record the transmission spectrum of the cavity. To this end the reference laser is adjusted to high-finesse polarization and locked to a TEM$_{11}$ transversal mode. The pump laser is adjusted to either high- or low-finesse polarization and slowly scanned across any transversal mode by ramping the voltage controlling the VCO (see Fig.~\ref{Fig1}). A recorded spectrum obtained from such a scan through a TEM$_{00}$ mode is shown in Fig.~\ref{Fig6}(a). Apparently the stability of the spectrometer is sufficient to resolve the $20~$kHz wide cavity resonance width. The reproducibility is such that the frequency separations between the transverse modes are always the same, even when the laser frequencies are changed by several THz.

A distinct feature of the transmission profile is its asymmetry. This has the following reason: While the probe laser is scanned through resonance, it locally heats the surfaces of the cavity mirrors. The mirrors slightly expand thus modifying the length and the free spectral range of the cavity. This shifts the resonance during the scan and leads to a hysteresis in the transmitted intensity. We have checked that the effect is suppressed for very weak probe laser powers. It is also much less pronounced for higher-order TEM modes, where the intensity is distributed over a larger area. This is seen in Fig.~\ref{Fig6}(b) exhibiting the profile of a TEM$_{04}$ mode.

Ideally the constructed spectrometer should be able to rule out any frequency jitter of the cavity modes, since the frequency of the cavity modes is instantaneously probed by the PDH servo and the information is passed via the PLL to the pump laser. In practice however, noise can enter the spectrometer via various channels. To estimate the quality of the PLL independently, we beat the free-running reference and the pump laser on a photodetector. The measured spectral width of the beat signal is narrower than 100~Hz, limited by the instrumental resolution of the detection system, provided a highly stable frequency synthesizer is used as local oscillator. Earlier \cite{Kruse03b}, the frequency deviations of the PDH-locked reference laser from the ring cavity mode have been found to be well below 1~kHz.

Unfortunately, the cavity itself appears to be subject to strong mechanical vibrations resonantly enhanced from environmental noise. At some particularly strong Fourier components the vibrations lead to frequency excursions of the locked reference laser as high as 500~MHz, as deduced from the noise in the intensity transmitted through an independent optical etalon. These frequency excursions tend to overstrain the PDH and PLL servo loops, and are at the origin of residual jitter observed in Fig.~\ref{Fig6}(a,b). The slope of the transmission profiles can be used as a discriminator for the frequency jitter. At integration times between 1~ms and 0.5~s, where the measured spectral density of the frequency fluctuations [see Fig.~\ref{Fig6}(c)] can be approximated by white noise, the pump laser linewidth (with respect to the resonant cavity mode) is estimated to 2~kHz \cite{Stewart54}.

\section{Atom-induced normal mode shifts}

The atoms located in the TEM$_{00}$ mode give rise to an additional refraction index for the light field propagating between the mirrors. Therefore, the optical path length is changed, leading to a shift in the resonance frequencies. This has also been described in \cite{Klinner06}. 
	\begin{figure}[h]
		\centerline{\scalebox{0.48}{\includegraphics{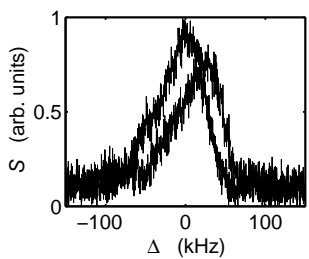}}}
		\caption{Transmission profiles of a TEM$_{00}$ mode in high finesse. The laser spectrometer is tuned $\Delta_a=(2\pi)~0.86~$THz to
			the red of the $D_2$ line. The right curve is taken with $N=1.5\times10^6$ atoms placed in the mode volume, the left curve is 
			taken without atoms. The maxima of both curves are separated by about $NU_1\simeq30~$kHz.}
		\label{Fig7}
	\end{figure}
The magnitude of this shift depends on the number of atoms interacting with the mode, so that we can calculate the atom number from the measured frequency shift. As explained in Section~\ref{SecSetup}, the single-photon light shift is $U_1=g_1^2/\Delta_a$. With $N$ atoms the mode is shifted by an amount $NU_1$. From the measurements shown in Fig.~\ref{Fig7} we estimate $NU_1\simeq30~$kHz and obtain the atom number $N\simeq1.9\times10^6$. This is in reasonable agreement with the atom number of about $2\times10^6$ measured by time-of-flight absorption imaging. The deviation is due to systematic errors arising from the asymmetric shape of the cavity transmission profiles.

\section{Conclusion}

In conclusion, we have demonstrated a method of probing the modes of a high-finesse cavity near an atomic resonance without perturbing an ultra-cold atomic cloud located within the cavity mode volume. In the present configuration, the red-detuned light of a reference laser is injected into a higher-order transversal cavity mode. Hence, the atoms are slightly attracted off center toward the intensity lobes of the mode. This could be avoided by tuning the laser to the blue of the atomic transition. In this way the atoms are repelled from the intensity maxima, which contributes to reducing radiation pressure.

The spectrometer presented in this work, together with a recently developed method to actively control backscattering from imperfections at the mirrors' surfaces \cite{Krenz07} will become important for future studies of collective dynamics in the regime of very weak coupling forces, where quantum effects are expected to play a role \cite{Piovella01b,Slama07b}.

\bigskip

This work has been supported by the Deutsche Forschungsgemeinschaft (DFG) under Contract No. Co~229/3-1.

\end{document}